\documentclass[reprint,
superscriptaddress,
amsmath,amssymb,aps,showkeys,showpacs,
twoside,final,secnumarabic,
nofootinbib]{revtex4-2}

\usepackage[paperwidth=205mm,paperheight=290mm,top=17mm,bottom=25mm,
inner=17mm,outer=17mm,
twoside]{geometry}

\usepackage{cmap} 
\usepackage[T1,T2A]{fontenc}
\usepackage[utf8]{inputenc}
\usepackage[russian,english]{babel}
\usepackage{color}
\usepackage{graphicx}
\usepackage{dcolumn}
\usepackage{bm} 
\usepackage[unicode=true,colorlinks=true,linkcolor=magenta, urlcolor=blue, citecolor = blue,breaklinks]{hyperref}
\usepackage{multirow}
\usepackage{url}
\usepackage{breakurl}
\DeclareGraphicsExtensions{.eps}

\newcount\issue
\newcount\Vol
\newcount\numb
\usepackage{fancyhdr} 
\def\Vol{\textbf{78}}
\def\numb{x}
\begin{document}

\title{Covariant Lagrangian   Cubic Interaction Vertices  \\  For Irreducible
Higher Spin Fields in Minkowski Backgrounds}

\def\addressa{address 1}
\def\addressb{address 2}

\author{\firstname{A.A.}~\surname{Reshetnyak}}
\email[E-mail: ]{reshet@tspu.edu.ru}
\affiliation{\addressa Center of Theoretical Physics, \\
Tomsk State Pedagogical University,\\
634061 Tomsk, Russia}

\affiliation{\addressb National Research Tomsk Polytechnic   University,\\
634050 Tomsk, Russia}


\begin{abstract}
We review   the application of  BRST and BRST-BV  approaches to construct the generic off-shell
local Lorentz covariant  cubic interaction vertices for irreducible
massless and massive higher integer spin fields (as the candidates for massive particles in the Dark Matter
problem) on $d$-dimensional Minkowski spaces.  It is shown that equivalence among  two  Lagrangian dynamics  for the
same cubically interacting fields with given masses and spins obtained by means of the approach with complete BRST, $Q$, operator and of one with
incomplete BRST, $Q_c$, operator in presence of  consistent off-shell holonomic (traceless) constraints can be uplifted from the equivalence of  the Lagrangians  for   free higher spin fields. We found that to get non-contradictory Lagrangians for irreducible   interacting higher-spin fields  within approach  with $Q_c$ operator, together  with off-shell   algebraic constraints   in addition to necessary condition of superconmmuting of $Q_c$ with appropriate holonomic constraints on the field and gauge parameter vectors, these constraints  should form Abelian superalgebra both with  BRST operator above and with operators of cubic vertices.
\end{abstract}

\pacs{Suggested PACS}\par 11.30.-j; 11.30.Cp; 11.10.Ef; 11.10.Kk; 11.15.~q

\maketitle
\thispagestyle{fancy}


\section{Introduction}\label{intro}

Interest to the higher spin (HS) field theory   is significant, due to its possibilities to provide a formulation of quantum gravity, then to open the door to physics beyond the Standard Model, by means of new elementary particles of matter and carriers of higher spin interactions, also massive HS fields  appear by legitimate candidates  to describe the problem of Dark Matter as an alternative to models of sterile and vector neutrinos.  A serious attention to HS fields
is due to its connection with String Field Theory on  constant curvature spaces, which operates with infinite sets of massive and massless fields of (half)integer  spins (for review and  references see, e.g. \cite{reviews3}, \cite{reviewsV},  \cite{Snowmass}).

A related principal problem in modern  high-energy theoretical physics  is the construction of  interactions  with  HS fields.
The structure of cubic and quartic vertices for various fields have been studied by   many authors within different approaches (see, e.g., the papers and references therein for cubic \cite{frame-like1}, \cite{BRST-BV3},  \cite{frame-like2},
 \cite{BRcub}, \cite{BRcubmass} and for quartic vertices  \cite{DT}).
The list of known results on cubic vertices on flat space-times, was given in terms of physical degrees of freedom in the light cone formalism in \cite{Metsaev0512}. In covariant metric form  a list of cubic vertices for reducible Poincare group representations  (compatible with \cite{Metsaev0512}) is contained in
\cite{BRST-BV3}, where the cubic vertices were deri\-ved with use of (constrained) BRST (Becchi-Rouet-Sto\-ra-Tyutin)  approach with incomplete, $Q_c$, BRST operator \cite{BRST-first} in presence of  off-shell holonomic constraints \cite{BRST-constr} imposed on Hilbert space vectors,  containing basic and auxiliary fields and gauge parameters.
 Recently, in  our papers \cite{BRcub}, \cite{BRcubmass}  the cubic vertices for massless and massive unconstrained HS fields on  spaces $\mathbb{R}^{1,d-1}$ were derived on a base of   BRST  approach with  $Q$  operator (see, e.g.  \cite{BResh1},  \cite{BKR}), which includes all constraints to describe irreducible representation (irrep) of given spin and mass within Noetherian-like deformation procedure of Lagrangian formulations (LFs) for free fields up to interacting  LF with preservation the number of physical degrees of freedom and irreducibility  for the  interacting  HS fields.
In result, an equivalence  among  LFs for the same HS fields obtained in the framework of approaches with  $Q$ and  $Q_c$  operators \cite{Reshetnyak_con} have been enlarged up to one for respective LFs of  interacting fields \cite{BRequiv} with resolving the peculiarities in \cite{BRST-BV3}.
\\
\indent
Equally powerful approach to get  gauge-invariant LF for free and interacting HS fields with $Q$ and  $Q_c$  BRST operators is realized within BRST-BV method   of constructing  BV  (Batalin--Vilkovisky) action  \cite{BV1}, which encodes the gauge algebra for the model with (ir)reducible  HS fields on the constant curvature  spaces, with complete, $Q$, BRST operator was suggested in \cite{BRSTBVResh} (and with  $Q_c$  BRST operator  \cite{BRST-constr}).  A construction of minimal BRST--BV actions is used both for the deformation procedure  and for finding the quantum gauge-fixed BRST-BV action.\\
\indent
The aim of the letter is to present the resumes of consistent construction  the   cubic  vertices  within
BRST and BRST-BV approaches with $Q$ and  $Q_c$   operators.\\
\indent
In the Section \ref{sec1:level1} we expose the general solution for  cubic vertices for unconstrained HS massless and massive fields.
A deformation for BRST-BV minimal action   with $Q$   operator is presented in Section~\ref{sec2:level1}. In Section~\ref{sec3:level1} we  cons\-truct the deformations of free LF and minimal BRST--BV action up to interacting ones for BRST\-(--BV) approaches  with incomplete $Q_c$ operator.\\
\indent
We use the definitions and notations  from
\cite{BRcub}, \cite{BRcubmass},  \cite{BRSTBVResh}  for a metric tensor $ \eta_{\mu\nu}$ and  $\epsilon(F)$, $(gh_H,gh_L, gh_{tot})(F)$,  $[x]$,
$(s)_{k}$, for the values of Grassmann parity and Hamiltonian, Lagrangian, total $gh_H+gh_L=gh_{tot}$   ghost numbers of a
 quantity $F$, the
integer part of
 real $x$, for  the  $k$-multiple  $
(s_1,s_2,...,s_k)$.

\section{\label{sec1:level1}Cubic vertices for unconstrained higher-spin fields}

Free massless   (massive) particle  with integer spin $s$ may be
described using the $\mathbb{R}$-valued totally symmetric field
$\phi_{\mu_1...\mu_s}(x)\equiv \phi_{\mu(s)}$ subject to the
equations
\begin{eqnarray}\label{irrepint}
    &&  \big(\partial^\nu\partial_\nu + m^2,\, \partial^{\mu_1},\, \eta^{\mu_1\mu_2}\big)\phi_{\mu(s)}  = (0,0,0)    \ \ \   \Longleftrightarrow  \  \\
     &&       \big(l_0,\, l_1,\, l_{11}, g_0 -d/2\big)|\phi\rangle  = (0,0,0,s)|\phi\rangle. \nonumber
\end{eqnarray}
The  vector $|\phi\rangle$ and the operators $l_0,\,
l_1,\, l_{11}, g_0 -d/2$ here  are defined in the Fock space $\mathcal{H}$
with the Grassmann-even oscillators $a_\mu, a^+_\nu$, ($[a_\mu, a^+_\nu]= - \eta_{\mu\nu}$):
\begin{eqnarray}\label{FVoper}
&&   |\phi\rangle  =  \sum_{s\geq 0}\frac{\imath^s}{s!}\phi^{\mu(s)}\prod_{i=1}^s a^+_{\mu_i}|0\rangle, \quad  \big(l_0,\, l_1,\, l_{11}, g_0\big)  \\
&&   = \big(\partial^\nu\partial_\nu+ m^2 ,\, - \imath a^\nu  \partial_\nu ,\, \frac{1}{2}a^\mu a_\mu ,  -\frac{1}{2}\big\{a^+_{\mu},\, a^{\mu}\big\}\big).\nonumber
\end{eqnarray}
The interacting LF within BRST approach with total $Q^{tot}=\sum_{i=1}^k Q^{(i)}$ BRST operator in cubic approximation is given with accuracy up to to be  linear in deformation parameter $g$, for $k$-copies of HS fields $(\phi^{(1)}_{\mu(s_1)},...,\phi^{(k)}_{\mu(s_k)})$ including in respective vectors  $|\phi^{(1)}\rangle, ...$ $|\phi^{(k)}\rangle$ of masses $(m)_k$ and spins $(s)_k$,   with own set of  oscillators $a^{(i)}_\mu, a^{+(i)}_\nu$
\begin{widetext}\begin{eqnarray}\label{S[n]}
  && S^{(m)_k}_{[1]|(s)_k}[(\chi)_k] \ = \  \sum_{i=1}^{k} \mathcal{S}^{m_i}_{0|s_i}[\chi^{(i)}]   +
g \sum_{1\leq i_1<i_2<i_3\leq k}  \int \prod_{j=1}^3 d\eta^{(i_j)}_0  \Big( {}_{s_{i_j}}\langle \chi^{(i_j)} K^{(i_j)}
  \big|  V^{(3)}\rangle^{(m)_{(i)_3}}_{(s)_{(i)_3}}+h.c. \Big)  , \\
   && \delta_{[1]} \big| \chi^{(i)} \rangle_{s_i}  =  Q^{(i)} \big| \Lambda^{(i)} \rangle_{s_i} -
g \sum_{1\leq i <i_1<i_2\leq k} \int \prod_{e=1}^{2} d\eta^{(i_e)}_0  \Big( {}_{s_{i_1}}\langle
\Lambda K^{({i_1})}\big|{}_{s_{i_2}}
   \langle \chi K^{({i_2})}\big| +(i_1 \leftrightarrow i_2)\Big)
\big|{V}{}^{(3)}\rangle^{(m)_3}_{(s)_{i(i)_2}} \label{cubgtr}\\
  && \delta_{[1]} \big| \Lambda^{(i)} \rangle_{s_i}  =  Q^{(i)} \big| \Lambda^{(i)1} \rangle_{s_i}
 -g  \sum_{1\leq i <i_1<i_2\leq k} \int \prod_{e=1}^{2} d\eta^{(i_e)}_0  \Big( {}_{s_{i_1}}\langle
(\Lambda^1 K)^{({i_1})}\big|{}_{s_{i_2}}
   \langle (\chi K)^{({i_2})}\big| +(i_1 \leftrightarrow i_2)\Big)
\big|{V}{}^{(3)}\rangle^{(m)_3}_{(s)_{i(i)_2}} \label{cubggtr}
\end{eqnarray}
\end{widetext}
(for $(\epsilon, gh_H)Q$=$(1,1)$; $(\epsilon, gh_H)\chi^{(i)}$=(0,0);$(\epsilon, gh_H)\Lambda^{(i)t}$ $=(t+1,-1-t)$, $t=0,1$).
These relations determine the first-stage reducible gauge theory with non-abelian gauge transformations (for $SU(N)$ gauge group, $k=N^2-1$) for the field vectors $| \chi^{(i)} \rangle_{s_i}$,  with gauge parameters of the zero $| \Lambda^{(i)} \rangle$  and first $| \Lambda^{(i)1} \rangle$  levels from total Hilbert space  $\mathcal{H}^{tot}=\oplus_i \mathcal{H}^{(i)}$ for  $\mathcal{H}^{(i)}$ related to $i$-th copy of fields. They, in turn, depend on additional auxiliary $b^{(i)}, b^{(i)+}$ ($d^{(i)}, d^{(i)+}$ for $m_i\ne 0$), and pairs of  ghost oscillators $\eta^{(i)}_0, \mathcal{P}^{(i)}_0$; $\eta^{(i)}_1, \mathcal{P}^{(i)+}_1$;   $\eta^{(i)}_{11}, \mathcal{P}^{(i)+}_{11}$; $\eta^{(i)+}_1, \mathcal{P}^{(i)}_1$;  $\eta^{(i)+}_{11}, \mathcal{P}^{(i)}_{11}$ \cite{BRcubmass}.
The  action,  equations of motion and spin for free HS field with  $(m_i, s_i)$ included in  $| \chi^{(i)} \rangle_{s_i}=  |\phi^{(i)}\rangle_{s_i} + \mathcal{O}(b^+,d^+, \eta^+)$
\begin{eqnarray}\label{Sclass}
  &&  \mathcal{S}^{m_i}_{0|s_i}[\chi^{(i)}] = \int d\eta^{(i)}_0   {}_{s_{i}}\langle (\chi  K)^{(i)}
   \big| Q^{(i)} \big| \chi^{(i)} \rangle_{s_{i}},\\
  && \nonumber    Q^{(i)} \big| \chi^{(i)} \rangle_{s_{i}} =0,  \ \sigma^{(i)} (\big| \chi^{(i)} \rangle, \big| \Lambda^{(i)} \rangle,\big|\Lambda^{(i)1} \rangle)_{s_{i}}=0,
  \end{eqnarray}
  (for $ \langle(\chi K)^{(i)}\big|  \equiv   \langle \chi^{(i)}\big| K^{(i)}$). Quantity $K^{(i)}$ is the non\--degenerate operator for preserving  reality of the action, and  complete  BRST,  spin  operators defined as
   \begin{eqnarray*}
&& {Q}^{(j)}\hspace{-0.1em} = \hspace{-0.1em} \big(\frac{1}{2}\eta_0l_0+\eta_1^{+}\check{l}{}_1+
{\frac{\imath}{2}}\eta_1^+\eta_1{\cal{}P}_0+ \widetilde{L}{}_{11} \eta^{+}_{11}\big)^{(j)}+h.c.,
\end{eqnarray*}
\vspace{-2ex}
\begin{eqnarray*}
&&\sigma^{(j)} = \big(G_0+\eta_1^+\mathcal{P}_{1}-\eta_1\mathcal{P}_{1}^+  + 2(\eta_{11}^+\mathcal{P}_{11} -\eta_{11}\mathcal{P}_{11}^+)\big)^{(j)}.
\end{eqnarray*}
Here $G_0, $, $\check{l}{}_1$, $\widetilde{L}{}^{(j)}_{11}$  are enlarged number particle operator $ G_0=g_0 + d^+d +2b^+b+ \frac{1}{2}+ h$, modified  divergence, BRST enlarged trace $\widetilde{L}{}^{(j)}_{11}= {l}{}^{(j)}_{11}$+$\mathcal{O}(b,d, \eta_1, \mathcal{P}_1)$ operators with theirs conjugated ones satisfying  the Lie algebra to be semidirect sum of $so(1,2)$ for $\widetilde{L}{}^{(j)}_{11}, \widetilde{L}{}^{(j)+}_{11}, G^{(j)}_0$ and isometry algebra  $l^{(j)}_0, \check{l}{}^{(j)}_1, \check{l}{}^{(j)+}_1$.

\begin{figure}[h]
{\footnotesize\begin{picture}(9,3)
\put(-35.5, -12.5){\includegraphics[scale=0.20]{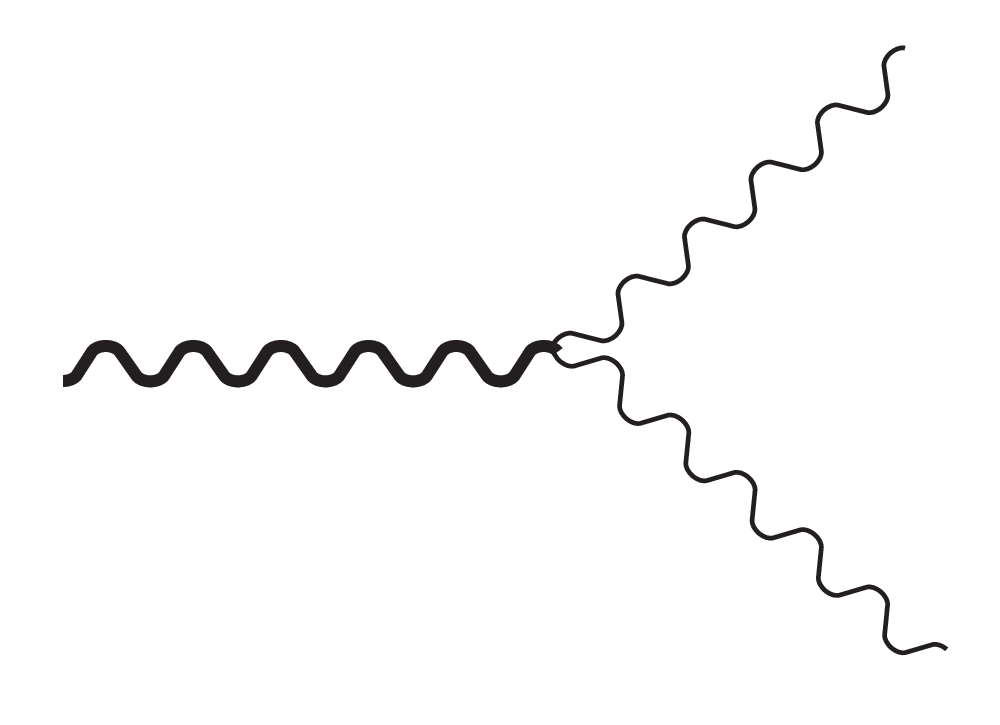}}
\put(-110.5, 20.5){$ |{V}{}^{(3)}\rangle^{(0,0,m)}_{(\lambda_1 ,\lambda_2,  s_3)}\ \ \equiv $ }
\put(-20.5, 30.5){$ \big(m, s_3\big)$ }
\put(55.5, 43.5){$\big(0,\lambda_1\big)$ }
\put(55.5, -0.5){$\big(0,\lambda_2\big)$}
\put(60.5, 20.5){$+ \ \ldots\ldots $ }
\end{picture}}
\caption{\label{m0m0m1}Cubic vertex    for  massive
$\phi^{(3)}_{\mu(s_3)}$  of spin $s_3$ and two massless fields $\phi^{(i)}_{\mu(\lambda_i)}$  of  helicities $\lambda_i$, $i=1,2$. The terms  in $"\ldots"$ correspond to the auxiliary fields  from $|\chi^{(i)}\rangle_{s_i}$}
\end{figure}

BRST operators are nilpotent on the Hilbert subspaces $\widehat{H}{}^{(i)}$ with definite spins: $\sigma^{(i)}\widehat{H}{}^{(i)}=0$,  and supercommute with $\sigma^{(i)}$.
The consistency  of interacting LF implies that the local  vertex $\big|{V}{}^{(3)}\rangle^{(m)_3}_{(s)_{(i)_3}}$  ($3$-vector in $\mathcal{H}^{tot}$),  satisfies to the \emph{deformation equations} \cite{BRcub}, \cite{BRcubmass}
\begin{equation}
\label{g1Lmod}
  \big(Q^{tot}, \sigma^{(i)}\big)
\big|{V}{}^{(3)}\rangle^{(m)_3}_{(s)_{(i)_3}} =0 .
\end{equation}
The general BRST-closed solutions with definite spins  were found  for some  triples from $k$ copies of  un\-con\-strained HS fields: massless \cite{BRcub}  with helicities $\lambda_i$;  2 massless of  helicities $\lambda_1, \lambda_2$ and  massive  $(m, s_3)$;  1 massless with $(0, \lambda_1)$  and  2 massive $(m_j, s_j)$, $j=2,3$  with coinciding   and  different masses \cite{BRcubmass}.\\
\indent
 E.g,,  the  vertex (given by Figure~\ref{m0m0m1}) for 2 massless\\
 \vspace{10ex}
\vspace{-3ex}
\noindent  and  massive  HS fields has the form
\begin{eqnarray}\label{genvertex}
  && |{V}{}^{(3)}\rangle^{m}_{(s)_3} \hspace{-0.15em} = \hspace{-0.7em}\sum_{(i,j_i)=(1,0)}^{3,[s_{i}/2]}\hspace{-0.7em} U^{(s_1)}_{j_1}U^{(s_2)}_{j_2}U^{(s_3)}_{j_3}|{V}{}^{M(3)}\rangle^m_{(s)_3-2(j)_3}, \\
  && {V}{}^{M(3)|m}_{{(s)_3-2(j)_3}} \hspace{-0.15em}  =   \hspace{-0.2em} \sum_{k} \hspace{-0.2em}\widetilde{\mathcal{L}}{}^{(3)}_{k}\prod_{i=1}^3 \mathcal{L}^{(ii+1)+}_{11|{\sigma_{i+2}}} , \  (s,J)  =   \sum_{i}\big(s_i , j_i\big). \nonumber
   \end{eqnarray}
(for $\sigma_{i} =   \frac{1}{2}(s-2J-k)-s_i  ,\ i=1,2; $,  $\sigma_{3}$ =   $\frac{1}{2}(s-2J+k)$ $-$ $s_3 $) and is $(3+1)$-parameters family to be  enumerated by the natural parameters $(j)_3$, and $k$ subject to the definite
relations \cite{BRcubmass}.   Among the representations for BRST-closed forms $\mathcal{L}^{(3)}_{k}$, $ \mathcal{L}^{(ii+1)+}_{11|{\sigma_{i+2}}}$ there exist ones to be polynomial and non-polynomial in oscillators. Explicitly mixed-trace operators are defined in \cite{BRcubmass} and the trace-deformed $ U^{(s_i)}_{j_i}$    polynomial forms (leading to new vertices  with less derivatives after gauging away the auxiliary fields from $|\chi^{(i)}\rangle_{s_i}$ as compared to ones in \cite{BRST-BV3})  look (for $L^{(3)}$   =     $\imath \widehat{\partial}{}^{(3)}_{\mu}a^{(i)\mu+}$ $-$ $\imath \widehat{\mathcal{P}}{}^{(3)}_0\eta_1^{(i)+}$) 
\begin{eqnarray}
&& \hspace{-0.1em} U^{(s_i)}_{j_i} \hspace{-0.1em} =\hspace{-0.1em}
(\widetilde{L}{}^{(i)+}_{11})^{(j_i-2)}\big\{(\widetilde{L}{}^{+(i)}_{11})^2
- j_i(j_i-1)\eta_{11}^{(i)+}\mathcal{P}_{11}^{(i)+}\big\}, \nonumber\\
\label{LrZ10}
  && \hspace{-0.1em} \mathcal{L}^{(3)}_{1}
   \hspace{-0.1em}= \hspace{-0.1em}  {L}^{(3)} - [\widetilde{L}{}^{(3)}_{11},{L}^{(3)}\}\frac{b^{(3)+}}{h^{(3)}}     , \\
\label{LrZ2m0}
  && \hspace{-0.1em} \widetilde{ \mathcal{L}}{}^{(3)}_{2}  =   (\mathcal{L}^{(3)})^2 -i\widehat{\mathcal{P}}{}^{(3)}_0\eta^{(3)+}_{11}-\widehat{l}_0^{(3)}  \frac{b^{(3)+}}{h^{(3)}}     ,   \\
  \label{LrZ2m}
  &&  \widetilde{\mathcal{L}}{}^{(3)}_{2k}  =  (\widetilde{\mathcal{L}}^{(3)}_{2})^k, \qquad \widetilde{ \mathcal{L}}{}^{(3)}_{2k-1} \ = \ (\widetilde{\mathcal{L}}^{(3)}_{2})^{k-1}\mathcal{L}^{(3)}_{1}.
\end{eqnarray}
The tensor representation for the vertices   with triples of HS fields $(m_i, s_i)$,   can be found with use of gauge fixing procedure for reducing a number of auxiliary fields  and  by calculation inner products \cite{BRcubmass}.

\section{\label{sec2:level1}BRST-BV approach for deformed  minimal BRST-BV action}

In turn, BRST--BV approach permits one to develop deformation procedure for LFs of the same HS fields in terms of minimal  BRST--BV   action. To formulate minimal BRST--BV action \cite{BV1}  with BRST $Q$ operator, we enlarge minimal configuration space $M^{(s)}_{\min}$  parameterizing by the fields $\Phi^{A} = (A^a, C^{\alpha_0}, C^{\alpha_1})$ (in condensed notations) with classical  $A^a=\phi_{\mu(s)}(x), ...,$, zeroth and first level ghost (to be in one-to-one correspondence with component tensors in  $|\Lambda^{(j)}\rangle$, $j=0,1$) fields by respective antifields $\Phi^*_{A} = (A^*_a, C^*_{\alpha_0}, C^*_{\alpha_1})$. In \cite{BRSTBVResh} it was shown that all field and antifield classical and ghost tensors can be nested into \emph{generalized field-antifield vector} $|\chi_{g}\rangle_s$ on augmented Hilbert space  $\mathcal{H}_g$.
It contains in addition to  \emph{generalized field} vector $|\chi_{\min}\rangle_s$ with $2^4$ component field vectors the  \emph{generalized  antifield} vector $|\chi^*_{\min}\rangle_s$ with $2^4$
 component antifield vectors  at independent ghost monomials $(\eta_0)^{n_0}(\eta_1^+)^{n_1}(\eta_{11}^+)^{n_{11}}(\mathcal{P}_1^+)^{p_1}(\mathcal{P}_{11}^+)^{p_{11}}$ for $n_{...}, p_{...}=0,1$
 \vspace{-1.5ex}
\begin{equation*}
 |\chi_{g}\rangle_s = |\chi_{\min}\rangle_s+|\chi^*_{\min}\rangle_s, \ \  |\chi^{(*)}_{\min}\rangle = |\chi^{(*)}\rangle + \sum_{p=0,1}|C^{(*)p}\rangle.
\end{equation*}
All  components in $|\chi_{g}\rangle_s$ are Grassmann-even with vanishing total ghost number values presented in the Table~\ref{table1} (for $p=0,1$)
\vspace{-2ex}
\begin{table}[htbp]
\begin{center}\caption{Grassmann parity and ghost numbers distributions for generalized (anti)field vectors}
\begin{tabular}{|c|cccccc|} \hline
                  &  $|C^p\rangle$ & $|\chi\rangle$  & $(\eta ,\mathcal{P})$   &    $|\chi^*\rangle$ & $|C^{*p}\rangle$ &  $|\chi^*_{\min}\rangle$ \\
                    \hline
                   $\epsilon$ & $0$&$0$& $(1,1)$  &  0 & 0& 0 \\
                   $gh_H$  & $-p-1$ &  0&(1,-1) & 1& $2+p$ &  - \\
                                      $gh_L$ & $p+1$&  0 & (0,0) & -1 & $-2-p$ &   - \\
                                                         $gh_{\mathrm{tot}}$& 0& 0& (1,-1)  &  0 & 0&  0  \\
                  \hline 
\end{tabular}\label{table1}.
\end{center}
\end{table}
The BRST-BV action ${S}^{m_i}_{0|s_i}\equiv S^{(s)}[\Phi_{\min}, \Phi^*_{\min}]$ with $(\epsilon, gh_{tot}){S}^{m_i}_{0|s_i}=0$,  for free HS field with $(m,s)$ has the same representation as the classical action   $\mathcal{S}^{m_i}_{0|s_i}[\chi^{(i)}]$ (\ref{Sclass}) under substitution: ${S}^{m_i}_{0|s_i}[|\chi_g\rangle_s] = \mathcal{S}^{m_i}_{0|s_i}\vert_{(\chi^{(i)}\to \chi^{(i)}_g)}$ \cite{BV1}.
It satisfies  to the classical  master-equation in terms of antibracket
\begin{equation}\label{mestand0}
  \big({S}_{0|s}, {S}_{0|s} \big)^{(s)}= 2{S}_{0|s} \frac{\overleftarrow{\delta}}{\delta\Phi^A}\frac{\overrightarrow{\delta}}{\delta\Phi^*_{A}}{S}_{0|s}=0
\end{equation}
The deformation procedure for construction an interacting gauge-invariant theory of $k$ copies  HS fields with $(m)_k,(s)_k)$  in BRST--BV method repeats
the one within BRST approach (\ref{S[n]}) for the cubic (for higher verices, see \cite{BRSTBVResh}) vertices  with  account of deformed gauge transformations (\ref{cubgtr}), (\ref{cubggtr}) are completely included into deformed  BRST-BV action, $S^{(m)_k}_{[1]|(s)_k}[(\chi_g)_k]$, obtained from  deformed classical action
when $(\chi)_k\to (\chi)_g$.

The cubically  deformed minimal action $\widetilde{S}_{[1]}$ =
$S^{(m)_k}_{[1]|(s)_k}[(\chi_g)_k]$ should satisfy master-equation with antibracket $\big(\cdot , \cdot\big)^{(s)_k}=\sum_i\big(\cdot , \cdot\big)^{s_i}$ defined in total field-antifield space coordinated by $\Phi^{Ai}, \Phi^*_{Ai}$, $i=1,...,k$
\begin{equation}\label{mestand}
  \big(\widetilde{S}_{[1]} , \widetilde{S}_{[1]}  \big)^{(s)_k}= 2\widetilde{S}_{[1]} \sum_{i}\frac{\overleftarrow{\delta}}{\delta\Phi^{Ai}}\frac{\overrightarrow{\delta}}{\delta\Phi^*_{Ai}}\widetilde{S}_{[1]} =0
\end{equation}
in total augmented Hilbert space  $\mathcal{H}^{tot}=\oplus_i \mathcal{H}^{(i)}_g$ with accuracy up to first degree in $g$.
Its solution in powers of $g$ leads for cubic approximation to the same equations (\ref{g1Lmod}) on the kernels of functional $\widetilde{S}_{[1]}$.

The minimal BRST-BV  action is invariant under deformed BRST-like transformations, $\delta_{[1]B}|\chi_{\min} \rangle_{s_i}= \mu \overrightarrow{s}_{[1]} |\chi_{\min} \rangle_{s_i}$,
 written in terms of Slavnov generator, $\overrightarrow{s}_{[1]}=\overrightarrow{s}_{0}+\overrightarrow{s}_{1}$, with nilpotent  $\mu$.
Suggested BRST--BV method with complete BRST operator and deformation  procedure for consistent interactions present the basic results of the section.
  The example of cubic vertex for  triple of massless HS fields $(0,\lambda)$, $(0,0)$, $(0,0)$ on $R^{1,d-1}$ was derived in  \cite{BRSTBVResh}.

\section{\label{sec3:level1}Consistent deformation  for constrained HS fields}

BRST and BRST--BV approaches with incomplete BRST operator $Q_c$ play significant role  in construction the LFs for interacting HS fields   due to
more restricted spectrum of auxiliary fields. For totally--symmetric case it is enough to have for $m_i=0$ ($m_i>0$) 2 additional  fields (sets) $\phi_{p|\mu(s-p)}$, $p=1,2$ known as triplet formalism which form restricted field vectors $|\chi^{(i)}_c\rangle_{s_i}$= $|\Phi^{(i)}\rangle_{s_i}-P_1^{(i)+}\sum_{p=0}^1\eta^{(i)+}_p +|\Phi^{(i)}_{p+1}\rangle_{s_i-p-1}$  not depending together with gauge parameters  $|\Lambda^{(i)}_c\rangle_{s_i}$ on oscillators related to traces: $b^+, \eta_{11}^+$ for $|\Phi\rangle\vert_{d^+=0} = |\phi\rangle$.

The principal result which provides consistent scopes for the (local) cubic vertices for irreducible interacting HS fields with $(m,s)_k$ within constrained  BRST  approach consists in the statement that  incomplete BRST, $Q_c^{tot}$, spin $(\sigma_c)_k$ and off-shell traceless constraints $L^{(i)}_{11}$ = $\widetilde{L}{}^{(i)}_{11}\vert_{b^+=h=0}$ operators and 3-vector of cubic vertices $ |{V}{}^{(3)}_c\rangle^{(m)_{(i)_3}}_{(s)_{(i)_3}}$ should obey the equations
\begin{eqnarray}
 &&   (Q_c^{tot})^2 \ = \  [ Q_c^{tot}, \,  L^{(i)}_{11}\} \ =\ [ Q_c^{tot}, \,  \sigma^{(i)}_{c}\}= 0, \nonumber \\
 \label{gencubBRSTc}
 &&  \left\{Q_c^{tot}, L^{(i)}_{11}  \Big)\right\}
\big|  V^{(3)}_c\rangle^{(m)_{(i)_3}}_{(s)_{(i)_3}} = (0,0).
\end{eqnarray}
The similar will be true for higher vertices. The respec\-ti\-ve form for  free $\sum_{j=1}^{k} {S}^{m_j}_{0|s_j}[\chi^{(j)}_c]$, interacting ac\-ti\-on $S^{(m)_k}_{[1]|(s)_k}[(\chi_c)_k]$  with cubic vertices are also descri\-bed by the relations (\ref{S[n]}), (\ref{cubgtr}), (\ref{Sclass}), for $K^{(i)}\equiv 1$ and vanishing "trace" oscillators.
The   $Q^{tot}_c$-closed and traceless  vertices $\big|  V^{(3)}_c\rangle$ for irreducible HS fields are related with respective reducible $Q^{tot}_c$-closed ones  $|{V}{}^{M(3)}_c\rangle$ in \cite{BRST-BV3} by means of trace projectors $P_{t|11}^{i_j}$, $t=0,1$  \cite{BRcubmass}, \cite{BRequiv}  as follows, $\big|  V^{(3)}_{c|t}\rangle = \prod_{j=1}^3P_{t|11}^{i_j}|{V}{}^{M(3)}_c\rangle $.


\section{Conclusion}

We described rhe essence and application of BRST and BRST--BV approaches with complete BRST  operator to consistently construct local cubic vertices for $k$-copies of interacting  HS fields on $d$-dimensional flat backgrounds with preservation the  Poincare group irreps for deformed gauge theory. The principal result reveals the fact  that it is possible to equivalently construct LFs for the same interacting HS fields  within approaches with complete and incomplete  BRST operators.  The necessary and sufficient conditions for construction cubic and higher vertices within    BRST approach with off-shell holonomic constraints  for irreducible interacting HS fields are established, which requires for the vertices to be traceless. The class of  local cubic vertices with less values of derivatives for unconstrained HS fields is found.

\begin{acknowledgments}
The author  wish to acknowledge the  Organizing Committee of the Lomonosov Conference, A.I. Studenikin for hospitality and wonderful scientific  spirit.
\end{acknowledgments}

\section*{FUNDING}
The work was  supported by  Ministry of
Education of Russian Federation, project No QZOY-2023-0003.


\nocite{*}


\end{document}